\documentclass[preprint,floatfix,aps,prd,showpacs,superscriptaddress,preprintnumbers]{revtex4-1}

\usepackage{graphicx}

\newcommand{\ba}{\begin{eqnarray*}}
\newcommand{\ea}{\end{eqnarray*}}
\newcommand{\bea}{\begin{eqnarray}}
\newcommand{\eea}{\end{eqnarray}}

\newcommand{\be}{\begin{equation}}
\newcommand{\ee}{\end{equation}}
\newcommand{\bd}{\begin{displaymath}}
\newcommand{\ed}{\end{displaymath}}

\newcommand{\yN}{y_{\rm{N}}}
\newcommand{\mev}{\mathrm{MeV}}

\newcommand{\fm}{\mathrm{fm}}

\def\eq#1{Eq.~(\ref{#1})}

\def\fig#1{Fig.~\ref{#1}}

\def\tab#1{Table~\ref{#1}}

\newcommand{\plotangle}{0}

\def\mcO{{\mathcal O}}

\def\mps{m_{\rm PS}}

\def\la{\langle}
\def\ra{\rangle}

\begin{document}

\title{Strangeness of the nucleon from Lattice Quantum Chromodynamics}

\author{Constantia Alexandrou}
\affiliation{Computation-based Science and Technology Research Center (CaSToRC), The Cyprus Institute,20 Constantinou Kavafi Street Nicosia 2121, Cyprus}
\affiliation{Department of Physics, University of Cyprus, P.O. Box 20537, 1678 Nicosia, Cyprus}
\author{Martha Constantinou}
\affiliation{Department of Physics, University of Cyprus, P.O. Box 20537, 1678 Nicosia, Cyprus}

\author{Simon Dinter}
\affiliation{NIC, DESY, Platanenallee 6, D-15738 Zeuthen, Germany}

\author{Vincent Drach}
\affiliation{NIC, DESY, Platanenallee 6, D-15738 Zeuthen, Germany}
\author{Kyriakos Hadjiyiannakou}
\affiliation{Department of Physics, University of Cyprus, P.O. Box 20537, 1678 Nicosia, Cyprus}

\author{Karl Jansen}
\affiliation{Department of Physics, University of Cyprus, P.O. Box 20537, 1678 Nicosia, Cyprus}
\affiliation{NIC, DESY, Platanenallee 6, D-15738 Zeuthen, Germany}

\author{Giannis Koutsou}
\affiliation{Computation-based Science and Technology Research Center (CaSToRC), The Cyprus Institute,20 Constantinou Kavafi Street Nicosia 2121, Cyprus}

\author{Alejandro Vaquero}
\affiliation{Computation-based Science and Technology Research Center (CaSToRC), The Cyprus Institute,20 Constantinou Kavafi Street Nicosia 2121, Cyprus}

\collaboration{ETM Collaboration}
\noaffiliation
\preprint{DESY 13-158,SFB/CPP-13-64}
\begin{abstract}

We present a non-perturbative calculation of the strangeness of the nucleon
$y_N$ within the framework of lattice QCD. This observable is known
to be an important cornerstone to interpret results from direct dark matter
detection experiments. We perform a lattice computation
for $\yN$ with an analysis of systematic effects originating from discretization,
finite size, chiral extrapolation and excited state effects
leading to the  value of $y_N= 0.173(50)$.
The rather large uncertainty of this value of $\yN$ is dominated by systematic
uncertainties which we were able to quantify in this work.
\end{abstract}


\maketitle

\section{Introduction}

The question of the exact composition of the nucleon, i.e. what are the
different quark contents of the proton and neutron,
is a long standing problem (see for instance \cite{Gasser:1982ap})
which can be addressed only through non perturbative methods.
In this paper, we will resort to lattice QCD techniques to address
the calculation of the  strange quark content of the nucleon which is important to know
not only for addressing the fundamental question of the nucleon composition,
but also, because it plays a most important role in the search for dark matter, as will be
discussed in more detail below.

A very useful measure for the strange quark content is the $\yN$-parameter,
\be\label{eq:def_yN}
\yN \equiv \frac{2\la N | \bar{s} s | N \ra }{\la N | \bar{u} u + \bar{d} d| N \ra },
\ee
where $u,d$ and $s$ denote respectively the up, down and strange quark fields.
We will refer to $\yN$ as the {\em strangeness of the nucleon} in the following.

As mentioned above, the $\yN$ parameter plays an important role in the context of dark matter searches.
Experiments which aim at a direct detection of dark
matter~\cite{Ahmed:2010wy,Bernabei:2008yi,Aalseth:2010vx,Sumner:2010zz,Jochum:2011zz}
are based on measuring the recoil energy of a nucleon hit by a dark matter candidate.
Even if in these processes a dark matter particle would not be
detected directly, such experiments allow
to provide bounds on the nucleon dark matter cross section
which can in turn be translated into constraints on models of New Physics.
In many supersymmetric scenarios~\cite{Ellis:2010kf} and in some Kaluza-Klein extensions of the standard model\cite{Servant:2002hb,Bertone:2010ww} the dark matter nucleon interaction is mediated through a Higgs boson.
In such a case
the theoretical expression of the spin independent scattering amplitude at zero momentum
transfer involves the $\yN$-parameter.
In fact, even rather small changes of the poorly known value of $\yN$
can be responsible for a variation of
one order of magnitude of the nucleon dark matter cross section. Having a better
determination of the $\yN$ parameter would thus provide better estimates
on the size of the cross-section or more reliable
constraints on dark matter models.

The $\yN$-parameter is related to the ratio of the pion-nucleon ($\sigma_{\pi N}$)
and the flavour non-singlet ($\sigma_0$) $\sigma$-terms, defined as
\begin{eqnarray}
\label{eq:sigma_terms}
\sigma_{\pi N} & \equiv &
m_l \langle N | \bar{u} u + \bar{d}d | N \rangle,\quad \sigma_{s}  \equiv m_s \langle N | \bar{s} s | N \rangle
\label{sigmapin} \\
\sigma_0 & \equiv & m_l \langle N | \bar{u} u + \bar{d}d -2 \bar{s} s| N \rangle
\label{sigm0}
\end{eqnarray}
where $m_l$  denotes the average up and down quark mass, and $m_s$ the strange quark mass
and where we also introduced the strange $\sigma$-term $\sigma_{s}$.
The $\sigma$-terms, $\sigma_{\pi N}$ and $\sigma_0$ can be estimated within
the framework of chiral effective field theories and using the
relation
\be
\yN = 1 - \frac{\sigma_0}{\sigma_{\pi N}} 
\label{relation}
\ee

also estimates of $\yN$ can be provided.

To be more specific,
the value of $\sigma_{\pi N}$ can be extracted from the pion nucleon
cross section data at an unphysical kinematics, known as the
Cheng-Dashen point.
Values for $\sigma_{\pi N }$ extracted in this way read
$\sigma_{\pi N }= 45 \pm 8~\mev$ from ref.~\cite{Gasser:1990ce} (GLS) and
$\sigma_{\pi N}=  64\pm 7~\mev$ from ref.~\cite{Pavan:2001wz} (GWU). A more recent
result has been obtained in ref.~\cite{Alarcon:2011zs} (AMO) which gives
$\sigma_{\pi N}=  59\pm 7~\mev$\cite{Alarcon:2011zs}.
The value of $\sigma_0$ can be computed analyzing the breaking of $SU(3)$ in the
spectrum of the octet of baryons. Following this strategy an estimate for this quantity is
given e.g. in \cite{Borasoy:1996bx} and reads  $\sigma^{\rm{I}}_{0}=  36\pm
7~\mev$. A more recent calculation using an improved method based on
Lorentz covariant chiral perturbation theory with explicit
decuplet-baryon resonance fields suggests  $\sigma^{\rm{II}}_{0}=
58\pm 8~\mev$\cite{Alarcon:2012nr}.

Using the aforementioned values for $\sigma_{\pi N}$ and  $\sigma^{I}_0$
we obtain the following phenomenological estimates of the $y_N$ parameter :
\be
y_N^{\rm{I,GLS}} = 0.20(21),~y_N^{\rm{I,GWU}} = 0.44(13),~y_N^{\rm{I,AMO}}=0.39(14).
\ee
Using $\sigma^{II}_0$ we obtain :
\be
y_N^{\rm{II,GLS}} = -0.29(29),~y_N^{\rm{II,GWU}} = 0.09(16),~y_N^{\rm{II,AMO}}=0.02(17).
\ee

Note that these values from effective field theory (EFT) and
phenomenology are affected by substantial errors
leading to correspondingly large uncertainties for the cross-section for dark matter
detection\footnote{Note that in \cite{Alarcon:2012nr}, the authors
obtain $\yN =  0.02(13)(10)$ and $\sigma_0$ around $60~\mev$}.

In this paper, we present a
first principle computation of the strangeness
of the nucleon using lattice QCD techniques. The difficulty of such
a computation has been for a long time that, due to the
appearance of dis-connected, singlet contributions, the error for
$\yN$ has been very large. Consequently, it has not been possible to obtain a precise
enough value which can be used for calculating the cross-section reliably,
see later in this manuscript for a discussion of various lattice computations.

In ref.~\cite{Dinter:2012tt} we were able to make a significant step
forward by using a setup of maximally twisted mass fermions
which avoids any mixing in the renormalization of
the $\sigma$-terms and hence $\yN$ does not need to be renormalized.
In addition, by employing special noise reduction techniques,
amenable for our setup, we could achieve a significant improvement
in the signal to noise ratio for $\yN$. The shortcoming of our result
in ref.~\cite{Dinter:2012tt} has been that, being a feasibility study only,
$\yN$ was obtained at only one value of the lattice spacing, a single finite
volume and only one quark mass.

Here we want to extend the calculation of ref.~\cite{Dinter:2012tt}
by using different lattice spacings, finite volumes and quark masses
such that we can probe effects of the discretization,
the finite volume and non-physical quark masses.
In addition, we now have available a high statistics analysis of
excited state effects which, as we will see below, are potentially
very dangerous for the computations of $\yN$.
Being able to address  the systematic uncertainties
appearing in a lattice calculation of $\yN$, we believe that our
computation can
provide  a reasonable estimate of the $\yN$ parameter based
on QCD alone and not resorting to effective field theories.

\section{Lattice QCD Calculation}

In our computation of $\yN$ we use gluon field configurations
generated by the
European Twisted Mass Collaboration (ETMC) \cite{Baron:2010bv} employing
maximally twisted mass fermions. In particular, the setup used here
includes a mass-degenerate light up and down
quark doublet as well as a strange-charm quark pair, a situation which
we refer to as the $N_f=2+1+1$ setup.
In our analysis
we have used two values of the lattice spacing, $a = 0.082~\fm$ and $a = 0.064~\fm$,
to examine lattice cut-off effects. We have a number of light
quark masses leading to
pseudo scalar meson masses $\mps$ covering the
range from $490~\mev$ to
$220~\mev$. This mass range allows us to perform
the chiral limit with $\mps$ approaching
the physical pion mass $m_{\pi}$.
We finally remark that we use a mixed action setup with Osterwalder-Seiler
quarks in the heavy quark sector which avoids any mixing due to the
iso-spin violation otherwise occurring in the twisted mass sea quark action.
This mixed action still enjoys the automatic $O(a)$-improvement of
twisted mass fermions.

The basic quantity needed for the evaluation of $\yN$ is the
ratio of correlation functions
\be\label{eq:def_R}
R(t,t_s) \equiv  \frac{ \sum_{\bf{x_{\rm{s}}},\bf{x}} \la  \bar{J}(x_{\rm{s}})  \Big(O_{\rm{s}}(x) -  \la  O_{\rm{s}}(x)\ra \Big)  J(0)\ra  }{\sum_{\bf{x_{\rm{s}}},\bf{x}} \la \bar{J}(x_{\rm{s}}) \Big( O_{\rm{l}}-  \la  O_{\rm{l}}(x) \ra \Big)   J (0) \ra},
\ee

where $J$ is an operator with quantum numbers of the nucleon and
$O_{\rm{l}} = \bar{u}u + \bar{d}d$ and $O_{\rm{s}} = 2 \bar{s}s$.
The calculation of the ratio of eq.~(\ref{eq:def_R}) is particularly challenging because of
very noisy contributions originating from dis-connected diagrams.
In \eq{eq:def_R},
$x = (t,{\bf{x}})$ and $x_s= (t_s,\bf{x_{\rm{s}}})$ denote the
Euclidean time and space coordinates.
We will refer to $t$ and $t_s$
as the source-operator separation and the source-sink separation, respectively.
We have shown in \cite{Dinter:2012tt} that $R(t,t_s)$ does not need to be renormalized since
no mixing in the renormalization pattern appears.
The ratio $R(t,t_s)$
has the following asymptotic behaviour:
\be\label{eq:limit_R}
R(t,t_s) = \yN + \mcO(e^{-\Delta M t }) + \mcO(e^{-\Delta M (t_{\rm{s}} - t) })
\ee
where we have denoted with $\Delta M$ the mass gap between the ground state and the
first excited state of the nucleon. Note that the two additional contributions
to $\yN$ are
non-vanishing as long as $t$ and $t_{\rm{s}}$ are finite. These two contributions
are a systematic effect inherent to any lattice calculation and will be
referred to as excited state contamination in the following.

More details on our setup and on the technique to evaluate \eq{eq:def_R}
can be found in \cite{Dinter:2012tt,Boucaud:2008xu} where we discuss in particular the crucial
points of our improved variance reduction technique and of the non-perturbative renormalization.

\section{Chiral extrapolation}

A crucial element in the determination of $\yN$ is the extrapolation
to the physical value of the pion mass. Ideally, to this end
chiral perturbation theory should be used. Let us therefore
shortly sketch, how the leading order chiral perturbation theory
behaviour of $\yN$ can be obtained.

Using the Feynman Hellman theorem, the sigma
terms can be related to the derivative of the nucleon mass with
respect to the light or strange quark masses:
\be
\sigma_{\pi N}  = m_l \frac{\partial m_N}{\partial m_l},\qquad
\sigma_{s} = m_s \frac{\partial m_N}{\partial m_s}
\label{eq:sigma_terms_FH}\, ,
\ee
and the $\yN$ parameter can be written as follows:
\be\label{eq:yn_FH}
\yN = 2  \frac{\partial m_N}{\partial m_s} \left( \frac{\partial \mps^2}{\partial
    m_l}\frac{\partial    m_N}{\partial \mps^2}\right)^{-1}
\ee
where we have neglected the strange quark mass dependence of $\mps^2$.

The nucleon mass dependence on the pion mass can be described by the
leading one-loop result in $SU(2)$ heavy baryon chiral perturbation theory
(HB$\chi$PT)~\cite{Gasser:1987rb} which reads:

\be
m _N(\mps)             = m^{(0)}_N - 4c^{(1)} \mps^2 - \frac{3
  g_A^2}{32\pi f_\pi^2} \mps^3 + \mcO(\mps^4)\; .
\label{eq:mN_Op3}
\ee
Here
$g_A$ is the axial coupling of the nucleon and $f_{\pi}$ is the
pion decay constant. We take the physical value  $g_A = 1.2695$ and
use the convention where the physical value of
$f_\pi$  is $92.4~\mev$. Note that the leading order expansion of the
the light sigma term reads $\sigma_{\pi N} = -4c^{(1)} \mps^2$
and thus $c^{(1)}$ has to be strictly negative to give
a positive value for $\sigma_{\pi N}$.

For the dependence of the nucleon strange sigma terms as a function of
the pseudo-scalar meson mass, we use the following ansatz:
\be
\la N| \bar{s} s |N\ra= d_0 + d_1 \mps^2 + \mcO(\mps^3)
\ee
with $d_0,d_1$ as fit coefficients.
Similar expressions can be
derived also from EFT, see for instance \cite{Chen:2002bz}.

Taking also the leading order chiral perturbation theory
expression of the pseudo-scalar
meson mass $m_{\rm{PS}}^2 = 2B m_{\rm{l}}$ into account,
the leading expression for the $\yN$ parameter reads:

\be\label{eq:yN_EFT}
\yN = \yN^{(0)} + \yN^{(1)} \mps + \mcO(\mps^2)
\ee
where
\begin{equation}
\yN^{(0)}= \frac{ d_0}{-4Bc^{(1)}},\qquad
\yN^{(1)} = \frac{9 d_0  g_A^2}{ 2 B \pi (4c^{(1)})^2 32 f_{\pi}^2}
\end{equation}

This simple leading order expression predicts
that $\yN$ is an increasing function of $\mps$. As we will see
in the following, this
contradicts the behaviour of our data for $\yN$. An interpretation of this
mismatch is that higher orders of chiral perturbation theory would be
needed to describe our results.
However, since this implies further free fit parameters
more data points than we have presently would be
necessary to be able to apply such higher order expressions.  In addition the validity of such a chiral expansion is questionable for pion masses above $300~\mev$ (see e.g \cite{Durr:2013goa}) and thus cannot be applied to describe most of the lattice data presented here.
This lack of being able to apply chiral perturbation theory
led us to use simple linear and quadratic fit ans\"atze
in the pseudo scalar mass, as will be discussed below.

\section{Results and systematic effects}

\begin{figure}
\includegraphics[width=0.7\textwidth,angle=\plotangle]{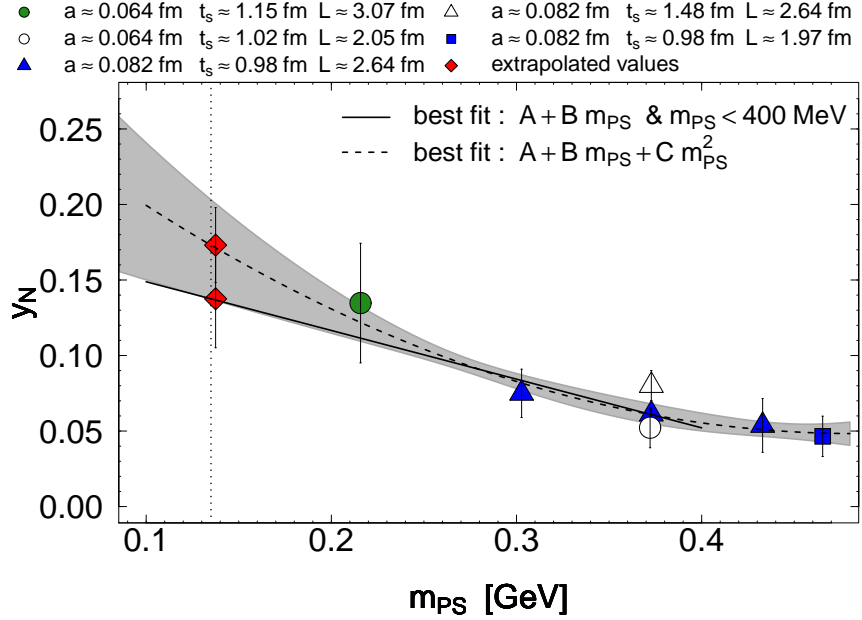}

\caption{Our results for $y_N$ as a function of
$\mps$. The values of the lattice spacing $a$, the linear extent of
the box $L$ and the source-sink separation $t_{{\rm s}}$ used here
are given in the legend. We extrapolate to the
physical value of the pion mass (marked by the vertical dotted line)
using linear (solid line) and
quadratic (dashed line) fits in $\mps$.
For the quadratic fit, we also show the corresponding error band.
Points represented by open symbols are only taken to
estimate systematic effects and are not included in our final analysis.}
\label{fig:yN}
\end{figure}

Our results for $\yN$ are shown in \fig{fig:yN}
as a function of $\mps$.
The chiral behaviour of the $\yN$ parameter is a difficult issue
and we are not aware of a direct computation for the quark mass
dependence of $\yN$ itself in the framework of EFT.
As discussed above,
our set of data points is insufficient to apply higher order chiral perturbation
theory.
It turned out that the number of fit parameters is too large
to obtain reliable fits and, with our data set, it was not possible
to disentangle different orders of the chiral expansion.
We therefore follow here the approach to
use simple polynomial fit ans\"atze for the pion mass dependence of $\yN$.

In the graph, we use different values
of the lattice spacing $a$, the physical linear extent of the box $L$ and
the source-sink separation $t_{\rm{s}}$.
We perform an extrapolation to the physical pion mass employing
a linear fit in $\mps$ (solid line) and a quadratic one
(dashed line). Note that only the points marked
by filled symbols are included in the fits and that only data for
$\mps<400~\mev$ are included in the linear fit.
Open symbols are solely used to demonstrate systematic effects which
will be discussed in more detail below.
The vertical dotted line in \fig{fig:yN} marks the physical
value of the pion mass.

In our Osterwalder-Seiler setup the value of the valence strange quark mass
has been  tuned  in order to match the Kaon mass obtained in the unitary setup.
In principle, also other matching conditions could be used leading
to different values of the valence strange quark mass.
By computing $\yN$ for valence strange quark
masses varying them by about 40\% we could not detect
any significant change in $\yN$ within our statistical error.
Hence, below we will not consider the tuning of the valence strange quark
mass as a source of systematic errors.

As mentioned above, the excited states contamination needs to be
scrutinized carefully in order to obtain reliable results.
This is particularly delicate in the  case of nucleon matrix elements
because the statistical error grows exponentially
when $t$ or $t_s$ are increased. Determining the asymptotic regime
in $t$ and $t_s$
where the two last terms of \eq{eq:limit_R} can be safely neglected
and a clear plateau behaviour appears
is thus often difficult given the typical statistics of
lattice calculations for nucleon observables.  We therefore performed
a detailed analysis of this effect on a single gauge ensemble
increasing by more than one order of
magnitude the statistics used.

\begin{figure}
\includegraphics[width=0.7\textwidth,angle=\plotangle]{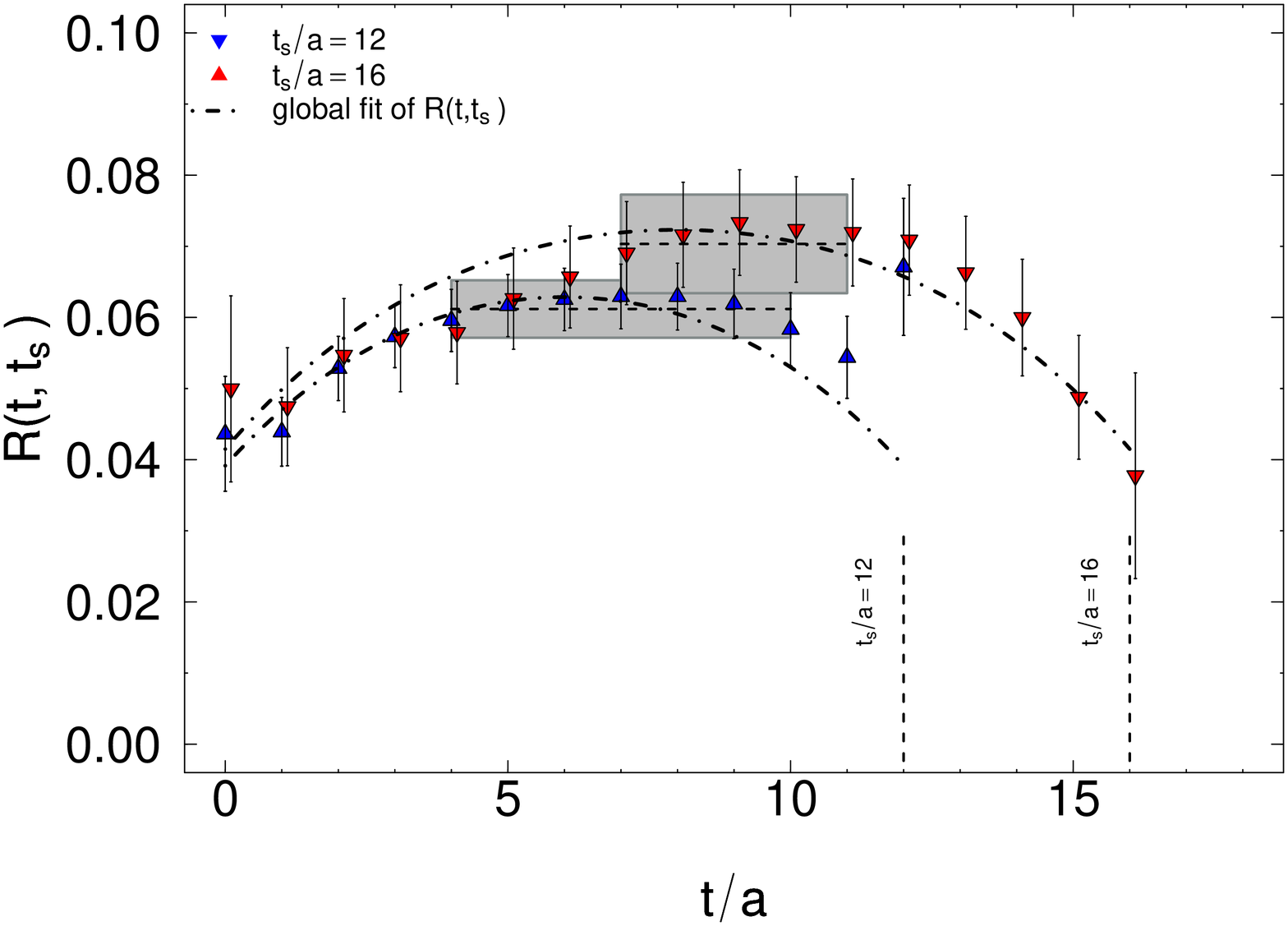}
\caption{$R(t,t_s)$ of \eq{eq:def_R} for two source-sink separations as
  a function of $t/a$. The
  source sink separations are indicated by vertical dotted lines. The
  best plateau fits together with their statistical error are
  represented by gray bands. The best fit curves of the global fit of
  $R(t,t_s)$ is shown by  black dotted curves.}
\label{fig:plateau}
\end{figure}

We computed the ratio $R(t,t_s)$ for
$t_s \sim 1.0~ \fm$ (filled  triangle in \fig{fig:yN}) and $t_s \sim 1.5 ~\fm$
(open  triangle in \fig{fig:yN}) keeping the value of $a=0.08~{\rm fm}$
and $L=2.6~{\rm fm}$ fixed.  We then performed several constant fits of the ratio $R(t,t_s)$
varying the fit interval $[t_{\rm{min}}/a,t_{\rm{max}}/a]$. We then
chose the longest plateau such that the fit on a restricted range  $[t_{\rm{min}}/a -1
,t_{\rm{max}}/a -1]$ change marginally compared to the one obtained
fitting on the range $[t_{\rm{min}}/a,t_{\rm{max}}]$. Since such a
fitting window can be found we conclude that the
systematic error introduced by the choice of a particular fitting
range is negligible. We summarize the fitting range dependence for
various source-sink separation in \tab{tab:excited_states}.
 Note that the plateaux have a good quality because of a large cancellation of the $t$ dependence  of the numerator and
denominator in the ratio $R(t,t_s)$.
In particular, we find $\yN=0.061(4)$ for $t_s=1.0~{\rm fm}$
and $\yN=0.080(10)$ for $t_s=1.5~ {\rm fm}$
which indicates a non-negligible excited states
contamination of about $\sim 32\%$. Note that we also performed
computations for intermediate source-sink separation of $1.14$ and
$1.31~\fm$. The results are summarized in \tab{tab:excited_states}
and as can be seen the latter source-sink separation is compatible with the result
obtained for source-sink separation of $\sim 1.5~\fm$. The effect of
different source-sink separations on
our data is also exhibited in \fig{fig:plateau} where we show  the ratio
$R(t,t_s)$ for two source-sink separation of $0.98~\fm$ and $1.31~\fm$. Note also that
in \cite{Dinter:2012tt} we obtained a result at $t_s = 1~\fm$ with a much lower
statistic ($\yN =0.082(16)$) which is fully compatible with the results
quoted here.  We thus consider as a conservative choice to use the difference between the two  values for $\yN$ obtained at $t_s=1.0~{\rm fm}$ and $t_s=1.5~{\rm fm}$ as an estimate of our systematic error originating from
excited states contamination assuming that this systematic
effect does not  depend strongly on the pseudo scalar meson
mass. As an additional check we performed global fits of the ratio
$R$ as a function of $t$ and $t_s$ for all the source-sink separations included in
\tab{tab:excited_states}.  The asymptotic behaviour of the ratio $R$ can be
written as follows assuming that only one state contributes:
\be
R(t,t_s) = \yN + Z (e^{-\delta m t} + e^{-\delta m (t_s - t)})
\ee
for a range of time parametrized by  $t \ge t_{\rm{cut}}$ and $t_s -t
\le t_{\rm{cut}}$ and where $\delta m$ is the difference of the energy of the first excited
state and the nucleon mass. The three coefficients $(\yN,Z,\delta m)$ are
treated as free parameters. The best fit values for $t_{\rm{cut}}/a=1$ reads 
\be
\yN = 0.091(7),\quad Z=0.047(5),\quad \text{and}~\delta m =  0.203(30)
\ee
with $\chi^2/{\rm{ndof}} = 23/53$. The corresponding curves are
shown in \fig{fig:plateau} for $t_s/a=12$ and $t_s/a=16$. Estimating $\yN$ in this way thus gives
a result compatible within statistical errors with the value obtained
for $t_s=18a$ quoted in  \tab{tab:excited_states}. The best fit value
for $\delta m \approx 490(70) \mev$ is larger than the mass of the
pion on this ensemble, as expected for a nucleon-pion 2 particle
state  at non zero
momentum.  We conclude that the difference between the results
obtained at  $t_s/a = 18$  and $t_s/a=12$ gives a reasonable estimate
of the systematic error due to the excited state contamination. 

Note that for the other gauge ensembles used in \fig{fig:yN}, 
we have used the same procedure to determine the fitting range
$[t_{\rm{min}}/a,t_{\rm{max}}/a]$ of the ratio $R(t,t_s)$ and also found a
marginal dependence of the results though with larger statistical errors.

\begin{table}[h]
\begin{center}
\begin{tabular}{c|c|c|c}
$t_s$ & $[t_{min}/a,t_{max}/a]$  & $y_N$ & $\chi^2/ndof$    \\
\hline
$12a \approx 0.98~\fm$ & $[3,9]$   & $0.061(4)$  & $1.3/6$ \\
$12a \approx 0.98~\fm$ & $[4,8]$   & $0.061(4)$  & $0.4/4$ \\
$14a \approx 1.15~\fm$ & $[3,11]$  & $0.063(5)$  & $2.33/8$ \\
$14a \approx 1.15~\fm$ & $[4,10]$  & $0.064(5)$  & $1.38/6$ \\
$16a \approx 1.31~\fm$ & $[6,10]$   & $0.070(7)$  & $0.72/4$ \\
$16a \approx 1.31~\fm$ & $[7,9]$   & $0.071(7)$  & $0.17/2$ \\
$18a \approx 1.48~\fm$ & $[5,13]$  & $0.080(10)$ & $2.2/8$ \\
$18a \approx 1.48~\fm$ & $[6,12]$  & $0.082(10)$ & $1.01/6$ \\
\end{tabular}
\caption{Dependence on the fitting window of $y_N$ for various source-sink separations. The
  fitting range and the $\chi^2/ndof$ are indicated.}
\label{tab:excited_states}
\end{center}
\end{table}

We have computed $\yN$ also at two different volumes
(filled  triangles and filled square in \fig{fig:yN}). However, we could not detect any
significant finite volume effects within the statistical errors and
thus  finite volume effects can be safely neglected.
We also show in \fig{fig:yN} results for $\yN$ for two different
lattice spacings (filled  triangle and empty circle in \fig{fig:yN}).
The two points are clearly
compatible, indicating that lattice discretization
effects are small. We took the difference between
the values of these two data points
as an estimate of the discretization errors.

In summary our final result reads:
\be
\yN = 0.173(29)(36)(19)(9)
\ee
where the central value is given by the quadratic fit,
 the first error is statistical, and the last three errors are our estimates of systematic
uncertainties, namely
the chiral extrapolation, the excited states contamination and the
discretization error, respectively. Note that the systematic errors
are partly substantially larger than the statistical one and
therefore dominate the
total error. Adding all errors in quadrature, we find
$\yN = 0.173(50)$.

\section{Discussion}

There are a number of lattice works that concentrate on the determination of
 $\sigma_s$   using direct and/or indirect computation (see for instance
 \cite{Toussaint:2009pz,Junnarkar:2013ac,Oksuzian:2012rzb,Gong:2013vja}  for $N_f=2+1$ results).
Other lattice works for $N_f=2+1$ provide indirect determinations of $y_N$
\cite{Durr:2011mp,Horsley:2011wr,Young:2009zb,Shanahan:2012wh} and our result
for $y_N$ is in agreement with these works.
We stress, however, that in our work we were
able to perform a comprehensive analysis of systematic uncertainties
covering lattice spacing and finite volume effects and, in particular, a
careful investigation of excited state contamination. Furthermore
computing directly the ratio of the matrix element $y_N$ allows us to avoid
any assumptions on the domain of validity of EFT relations which is based on
$SU(2)$ or $SU(3)$  $HB\chi PT$ expansion and sometimes known only at
leading order accuracy.

\begin{figure}[htb]
\begin{center}
\includegraphics[width=0.7\textwidth,angle=\plotangle]{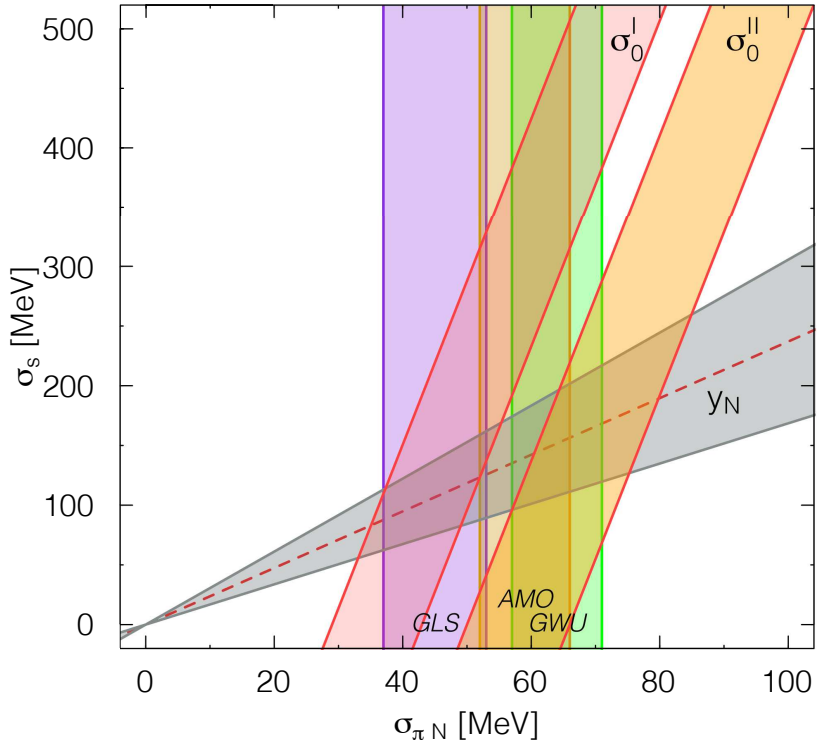}
\end{center}
\caption{Constraints on $\sigma_s$  obtained from our determination of
$\yN$. The phenomenological determination of $\sigma_{\pi N}$
are represented by colored band as obtained from
\cite{Gasser:1990ce} (GLS),
\cite{Pavan:2001wz} (GWU) and \cite{Alarcon:2011zs} (AMO). We also
show the constraint provided by the  estimates  $\sigma^{\rm{I}}_0$
\cite{Borasoy:1996bx} and $\sigma^{\rm{II}}_0$
\cite{Alarcon:2012nr}. As can be seen the value of $\yN$
can constrain the value of $\sigma_s$ to be smaller than about
$250~\mev$. }\label{fig:constraints}
\end{figure}

As  in  \cite{Young:2013nn}, we show in \fig{fig:constraints} the
$(\sigma_{\pi N},\sigma_s)$ plane together with vertical colored
bands that represent the phenomenological determinations and the
corresponding uncertainties of $\sigma_{\pi N} $
mentioned in the introduction. In order to put further
constraints on $\sigma_s$, we use the following relations $\sigma_s= \frac{1}{2} \frac{m_s}{m_l} \left( \sigma_{\pi N} - \sigma_0 \right) =  \yN \frac{1}{2}\frac{m_s}{m_l} \sigma_{\pi N} $
together with the ratio of the quark masses taken from the FLAG
group \cite{Colangelo:2010et}.
A first constraint derives from using
the phenomenological determination of $\sigma_0$ (indicated
by $\sigma_0$ in \fig{fig:constraints}).
The work performed here provide s a constraint through our
direct computation of $\yN$ (gray contour) which includes the estimate of
both statistical and systematic errors. As can be seen the result
constraints the strange $\sigma$-term and suggests an
upper bound of $\approx 250~\mev$ for $\sigma_s$.

\section{Conclusions and Outlook}

In this work, we have
performed a direct computation of the strangeness of the
nucleon $\yN$, including light, strange and charm sea quarks
with an emphasis on the study of systematic
effects. Using maximally twisted mass fermions
which allow for an efficient noise reduction technique and
which avoids mixing under renormalization we have obtained
$\yN=0.173(50)$.
Our result for $\yN$ is
compatible with previous determinations \cite{Bali:2011ks,Durr:2011mp,Horsley:2011wr,Young:2009zb,Shanahan:2012wh}
but includes an analysis of systematic errors originating from discretization, chiral extrapolation and excited states uncertainties.
It is worth pointing out that we find a rather low value of
the strange $\sigma-$term with corresponding consequences for the nucleon-dark matter
cross section.

One important conclusion of our work is that
the error we obtain for
$\yN$ is dominated by systematic uncertainties, in particular
the chiral extrapolation and excited state contamination which
cannot be neglected.
While the error from the chiral extrapolation can be avoided
in future calculations which are performed at or very
close to the physical value of the pion mass, the excited state
contamination must be carefully assessed.
Thus, future lattice evaluation of the $\yN$ parameter using the physical value of the light quark mass will avoid the systematic error due to the chiral extrapolation, but they will still have to address excited state contamination, as this study has demonstrated.

\section*{Acknowledgments}

We thank our fellow members of ETMC for their constant collaboration.
We are grateful to the John von Neumann Institute for Computing (NIC),
the J{\"u}lich Supercomputing Center and the DESY Zeuthen Computing
Center for their computing resources and support.  This work has been
supported in part by the DFG Sonderforschungsbereich/Transregio
SFB/TR9-03,
by the Cyprus Research Promotion
Foundation under contracts KY-$\Gamma$/0310/02/,
and the Research Executive Agency of the European Union under Grant Agreement number PITN-GA-2009-238353 (ITN STRONGnet).
K. J. was supported in part by the Cyprus Research Promotion
Foundation under contract
$\Pi$PO$\Sigma$E$\Lambda$KY$\Sigma$H/EM$\Pi$EIPO$\Sigma$/0311/16.

\bibliography{letter}
\bibliographystyle{h-physrev}

\end{document}